\documentclass[prd,twocolumn,showpacs,showkeys,preprintnumbers,amsmath,amssymb,floatfix]{revtex4}

\usepackage[utf8]{inputenc}
\usepackage{hyperref}

\usepackage{dcolumn}
\usepackage{bm}
\usepackage{graphicx}
\usepackage{subfigure}

\def\be{\begin{equation}}
\def\ee{\end{equation}}
\def\bestar{\begin{equation*}}
\def\eestar{\end{equation*}}

\def\t#1{\tilde #1}
\def\p{\partial}

\begin{document}

\title{Non-commutative Einstein-Proca Space-time}

\author{Angélica González}
\email{gombstar@xanum.uam.mx}
\affiliation{Departamento de Física, Universidad Autónoma Metropolitana-Iztapalapa\\
                   A.P. 55-534, C.P. 09340, M\'exico D.F., México}

\author{Román Linares}
\email{lirr@xanum.uam.mx}
\affiliation{Departamento de Física, Universidad Autónoma Metropolitana-Iztapalapa\\
                   A.P. 55-534, C.P. 09340, M\'exico D.F., México}

\author{Marco Maceda}
\email{mmac@xanum.uam.mx}
\affiliation{Departamento de Física, Universidad Autónoma Metropolitana-Iztapalapa\\
                   A.P. 55-534, C.P. 09340, M\'exico D.F., México}
                   
\author{Oscar Sánchez-Santos}
\email{oscarsanbuzz@yahoo.com.mx}
\affiliation{Departamento de Física, Universidad Autónoma Metropolitana-Iztapalapa\\
                   A.P. 55-534, C.P. 09340, M\'exico D.F., México}

\date{\today}

\begin{abstract}
In this work we present a deformed model of Einstein-Proca space-time based on the replacement of point-like sources by non-commutative smeared distributions. We discuss the solutions to the set of non-commutative Einstein-Proca equations thus obtained, with emphasis on the issue of singularities and horizons.
\end{abstract}


\maketitle

\section{Introduction}
\label{intro}

Non-commutativity in the classical formulation of 4-dimensional General Relativity can be implemented in several ways. The most frequently used is based in the replacement of the ordinary product of functions by the Wigner-Weyl-Groenewold-Moyal $\star$-product~\cite{ Groenewold:1946kp,Moyal:1949sk}
\be
(f \star g)(\bm x) = e^{\frac i2 \theta^{\mu\nu} \p_\mu \p_\nu} f(\bm x) g(\bm y) |_{\bm y \to \bm x}.
\ee
In this approach, instead of using non-commutative operators for the coordinates, one uses commutative coordinates variables and a modified point-wise product of functions of them. In most of the models, the elements $\theta^{0i}, \, i = 0, \dots 3$, are taken to vanish and only non-commutativity among spatial coordinates is considered. 

Although this prescription is easy to implement, theories based on it are sometimes difficult to solve explicitly due to the fact that a general $\star$-product involves an infinite number of derivatives, being thus non-local. This feature forces in some situations to resort to a perturbative treatment as perhaps the only way to gain some insight into the problem at hand. This is properly justified in physical scenarios where the non-commutative parameter is thought to be small, recent bounds on it are of the order of $(\mbox{10 Tev})^{-2}$, however it is always interesting to ask what happens if this is not the case, for example on a quantum gravity theory.

We should recall that one of the goals of non-commutativity when used in a classical theory is the regularisation of singularities, namely, classical divergences are expected to be removed by the introduction of a non-commutative structure to which is associated a fundamental length. The first model where this idea was put into practice is due to Snyder~\cite{Snyder:1946qz} with his formulation of Special Relativity aiming to eliminate the infinities appearing in the early stages of Quantum Electrodynamics~\cite{Snyder:1947nq}.

It has been pointed out that a non-commutative structure may violate Lorentz invariance if the time coordinate is involved. In order to remedy this situation, a proposal to implement non-commutativity such that Lorentz invariance is maintained has been formulated a few years ago in the context of non-commutative quantum field theory (NCQFT) by Smailagic and Spallucci~\cite{Smailagic:2003rp,Smailagic:2004yy} using coherent states. Among the several interesting results of this approach, it has been noticed that non-commutativity effectively replaces point-like behaviour in classical equations, such as Dirac delta functions, by smeared sources described by Gaussian functions. This is in accordance with the point of view that space-time, when equipped with a non-commutative structure, becomes fuzzy and points are not longer localised. 

More recently, the above result has been applied to the formulation of non-commutative gravitational models~\cite{Nicolini:2005vd,Nicolini:2008aj,Tejeiro:2010gu,Liang:2012vx,Rahaman:2013gw}. In analogy with NCQFT, sources of matter and charge are given by Gaussian distributions that in the commutative limit reproduce the classical singular behaviour. With these models, classical issues such as the presence of horizons in black hole solutions have been reexamined.

In the following we will use this strategy to build a deformed Einstein-Proca system. In his original motivation, Proca~\cite{Proca:1936} aimed for an equation describing particles with positive energy states, having the possibility of both signs of charge and with non-vanishing spin. As it is well-known, Proca Lagrangian extends naturally electromagnetism to the massive photon case and thus the Proca vector field has associated three polarisations states, there are a characteristic length for electromagnetic interaction that depends inversely on the mass of the photon, conservation of charge is preserved and it is relativistic invariant~\cite{Pauli:1941zz,Tu:2005ge}. 

The use of Proca's equations in physics have been diverse and important, for example spin-1 mesons are described by them. More recently, Proca-like Lagrangians have been used in the analysis of renormalizability of massive Abelian gauge theories, Lorentz violation within the photon sector of the Standard Model and also in the formulation of Abelian gauge theories in Very Special Relativity~\cite{Casana:2008nw,Cheon:2009zx}. When considering coupling to gravitational field, the Einstein-Proca system leads to interesting results on the issue of black holes and their singularities~\cite{Vuille:2002qz} and this line of research is what we want to pursue.

The present work is organised as follows: in Section~\ref{secc:2} we describe the non-commutative structure of the Einstein-Proca model we consider. Then, in Section~\ref{secc:3}, the deformed field equations of the model are solved. A class of regular solutions at the origin is then given in Section~\ref{secc:4} and the strong non-commutative regime, where more explicit calculations can be done, is discussed. Finally, we end with some general remarks of the results obtained in the Conclusions.

\section{Non-commutative Einstein-Proca Space-time}
\label{secc:2}
	
The non-commutative Einstein-Proca model is obtained by considering the field equations
\be
G^\mu{}_\nu = \kappa [ (T_{nc})^\mu{}_\nu |_{matter} + T^\mu{}_\nu |_{el} ],
\ee
for the spherically symmetric metric 
\be
ds^2 = e^{\nu(r)} dt^2 - e^{\lambda(r)} dr^2 - r^2 (d\theta^2 + \sin^2 \theta d\phi^2).
\ee
In this case, the energy-momentum tensors in the field equations have the following structure: first, the non-commutative matter tensor, $(T_{nc})^\mu{}_\nu |_{matter}$, is given by
\be
(T_{nc})^\mu{}_{\nu} = diag (h_1, h_1, h_3, h_3), \quad h_3 := h_1 + \frac 12 x \partial_x h_1, 
\ee
where
\be
h_1 := -\rho_m = - \frac M{(4\pi\theta)^{3/2}} e^{-r^2/4\theta}.
\ee
Here $\rho_m$ represents a non-commutative diffused distribution of matter which in the commutative limit $\theta \to 0$ becomes a point-like source of mass $M$ located at the origin. It can be verified straightforwardly that this form of the energy-momentum tensor satisfies $\nabla_\mu (T_{nc})^{\mu\nu} = 0$ due to the definition of the function $h_3$ in terms of $h_1$.

For the electromagnetic part, we have~\cite{Vuille:2002qz}
\be
T^\mu{}_\nu |_{el}  = 2\alpha F_\mu{}^\delta F_{\nu\delta} + \beta A_\mu A_\nu - \frac 12 g_{\mu\nu} (\alpha F_{\rho\sigma} F^{\rho\sigma} + \beta A_\rho A^\rho),
\ee
where $A_ \mu$ is the four-electromagnetic potencial and $F_{\mu\nu} := \p_\mu A_\nu - \p_\nu A_\mu$ is the electromagnetic tensor. This expression comes from the Proca-like Lagrangian density
\be
{\cal L} = \sqrt{-g} \left( \alpha F_{\mu\nu} F^{\mu\nu} + \beta A_\mu A^\mu \right).
\ee
Strictly speaking, the Proca Lagrangian corresponds to the choice 
\be
\frac \beta{2\alpha} = - \mu^2,
\ee
where $\mu$ represents the mass of the gauge field.

The electromagnetic fields are determined by the conservation laws
\be
\frac 1{\sqrt{-g}} \partial_\mu \left( \sqrt{-g} F^{\mu\nu} \right) - \frac \beta{2\alpha} A^\nu = \frac 1{4\alpha} J_{nc}^\nu,
\ee
where now we have a non-commutative four-current $J_{nc}^\nu$ in this equation of the form
\be
J_{nc}^\mu = (g_1, \vec 0), \qquad g_1 := \rho_e = \frac Q {(4\pi\theta)^{3/2}} e^{-r^2/4\theta}.
\ee
Here $\rho_e$ represents the diffused density charge of the source at the origin which in the commutative limit becomes point-like, in a similar fashion as for the mass density in the non-commutative energy-momentum tensor.

\subsection{Non-commutative field and conservation equations}

With the above definitions, we now proceed along the same lines as Vuille et al~\cite{Vuille:2002qz}. First, we set $A^\mu = (A_0, 0, 0 ,0 )$ and use in the field equations and conservation laws the following parametrisations 
\be
\alpha = - \frac 12 \epsilon_0, \quad \beta = \mu^2 \epsilon_0, \quad s := \frac {q\mu}{\epsilon_0},
\ee
together with the dimensionless variables $r$ and $u$ for the position and electric field defined by the relations
\be
x := \mu r, \quad A_0 =: s u.
\ee
After this, we make then a series expansion of all functions, but $h_1, h_3$ and $g_1$, on the parameter 
\be
\epsilon = \kappa \frac {q^2 \mu^2}{\epsilon_0},
\ee
which on physical grounds will be small~\cite{Vuille:2002qz}. In this way we arrived to the following set of equations of order zero on the parameter $\epsilon$ 
\begin{eqnarray}
-\frac 1{x^2} + \frac {e^{-\nu_0}}{x^2} - \frac {e^{-\nu_0} \kappa h_1}{\mu^2} - \frac {\nu_0^\prime}x = 0, 
\nonumber \\[4pt]
- \frac {\kappa h_3 e^{-\nu_0}}{\mu^2} - \frac {\nu_0^\prime}x - \frac 12 \nu_0^{\prime\, 2} - \frac 12 \nu_0^{\prime\prime} = 0,
\nonumber \\[4pt]
- \frac {g_1}{2s\mu^2 \epsilon_0} - e^{-\nu_0} u_0 + 2 \frac {u_0^\prime}x + u_0^{\prime\prime} = 0.
\label{PNC13}
\end{eqnarray}
These equations are simply those associated to the gravitational field of a particle sitting at the origin with both fuzzy mass and charge; their solution are straightforward to obtain. 

The corresponding field equations to first order on $\epsilon$ are more involved
\begin{widetext}
\begin{eqnarray}
-\frac 12 e^{-2\nu_0} u_0^2 + \frac {e^{-\nu_0} \lambda_1}{x^2} - \frac {\kappa h_1 e^{-\nu_0} \lambda_1}{\mu^2} - \frac 12 e^{-\nu_0} u_0^{\prime \, 2} + \frac {\lambda_1^\prime}x  = 0,
\nonumber \\[4pt]
\frac 12 e^{-2\nu_0} u_0^2 + \frac {e^{-\nu_0} \lambda_1}{x^2} - \frac {\kappa h_1 e^{-\nu_0} \lambda_1}{\mu^2} - \frac 12 e^{-\nu_0} u_0^{\prime \, 2} - \frac {\nu_1^\prime}x  = 0,
\nonumber \\[4pt]
\frac 12 e^{-2\nu_0} u_0^2 - \frac {\kappa h_3 e^{-\nu_0} \lambda_1}{\mu^2} +  \frac 12 e^{-\nu_0} u_0^{\prime \, 2} + \frac {\lambda_1^\prime}{2x} + \frac 14 \lambda_1^\prime \nu_0^\prime - \frac {\nu_1^\prime}{2x} - \frac 34 \nu_1^\prime \nu_0^\prime - \frac 12 \nu_1^{\prime\prime} = 0,
\nonumber \\[4pt]
- e^{-\nu_0} u_1 - \frac {g_1 \lambda_1}{2s\mu^2 \epsilon_0} - e^{-\nu_0} u_0 \lambda_1 - - \frac {g_1 \nu_1}{2s\mu^2 \epsilon_0} + 2 \frac {u_1^\prime}x - \frac 12 u_0^\prime \lambda_1^\prime - \frac 12 u_0^\prime \nu_1^\prime + u_1^{\prime\prime} = 0.
\label{PNC14}
\end{eqnarray}
\end{widetext}
In the next section we discuss the solutions of all these equations. 

\section{Solving the field equations}
\label{secc:3}

First we solve the set of equations to zero order as a check for our calculations and because these solutions will be used in the equations to first order. We multiply then with $-x^2e^{\nu_0}$ the first line in Eq.~(\ref{PNC13}) to get the following expression 
\begin{equation}
x e^{\nu_0}\nu^\prime_0+e^{\nu_0} = 1 - \frac{\kappa x^2h_1}{\mu^2}.
\label{PNC15}
\end{equation} 
The terms on the left hand side form a total derivative, thus we have
\begin{equation}
\frac {d(xe^{\nu_0})}{dx}=1 - \frac {\kappa x^2 h_1}{\mu^2},
\label{PNC16}
\end{equation}
and performing the integration on $x$ leads to 
\be
e^{\nu_0} = 1 + \frac bx - \frac \kappa{\mu^2} \int_0^x dz\, z^2 h_1(z),  
\ee
where $b$ is an integration constant. Regular solutions at $x = 0$ are posible if we set $b = 0$. Using the explicit expression for $h_1$ in this case, we obtain
\begin{equation}
e^{\nu_0} = 1 - \frac {\kappa\mu M}{2\pi^{3/2}} \frac 1x \gamma \left( \frac 32, \frac {x^2}{4\mu^2\theta} \right),
\label{PNC17}
\end{equation}
where $\gamma$ is the lower incomplete gamma function
\begin{equation}
\gamma\left( n, z \right) := \int_0^z dt\, t^{n-1/2}e^{-t}.
\label{PNC18}
\end{equation}
This is the result previously obtained by Nicolini~\cite{Nicolini:2005vd}.

The second line in Eqs.~(\ref{PNC13}) can be seen to be equivalent to the previous equation since we can rewrite it as 
\begin{equation}\label{PNC19}
\frac {d^2(xe^{\nu_0})}{dx^2} = - \frac{2\kappa}{\mu^2} x h_3.
\end{equation}
Integration of this equation with respect to $x$ gives the same result as in Eq.~(\ref{PNC17}) due to the identity
\be
\int^x dz \, z h_3 (z) = \frac 12 x^2 h_1(x),
\ee
which can be easily verified using the definition of $h_3$ in terms of $h_1$ and an integration by parts. In this way the system to zero order is solved completely for the metric functions. 

The equation for the dimensionless electromagnetic potential $u_0$ is given in the third line of Eqs.~(\ref{PNC13}). In order to solve it, we make the following considerations: first, along the same line of thought as in~\cite{Vuille:2002qz}, we take the point of view that the mass term in the solution Eq.~(\ref{PNC17}) is a correction to the Minkowski metric, thus we write
\begin{equation}
e^{\nu_0(x)}=1+\epsilon(x),
\label{PNC20}
\end{equation}
where 
\begin{equation}
\epsilon(x) = - \frac {\kappa\mu M}{2\pi^{3/2}} \frac 1x \gamma \left( \frac 32, \frac {x^2}{4\theta\mu^2} \right),
\label{PNC21}
\end{equation}
is accordingly a function representing small perturbations around flat space-time. Consequently, we can safely set $e^{\nu_0(x)}=1$ in the last of the zero-order field equations. In this way, the equation for $u_0$ in Eqs.~(\ref{PNC13}) reduces to 
\begin{equation}
- \frac {g_1}{2s\mu^2 \epsilon_0} - u_0 + 2 \frac {u_0^\prime}x + u_0^{\prime\prime} = 0.
\label{PNC22}
\end{equation}
The homogeneous solutions of this differential equation are $y_0 = e^{-x}/x$ and $y_1 = e^x/x$. Then, using the variation of parameters algorithm, we can write down the full solution in the form
\begin{eqnarray}
u_0 &=& c_0 y_0 + c_1 y_1 - y_0 \int_0^x dz\, \frac {y_1(z) \t g_1(z)}{W(y_0, y_1)(z)} 
\nonumber \\[4pt]
&&+ y_1 \int_0^x dz\, \frac {y_0(z) \t g_1(z)}{W(y_0, y_1)(z)},
\label{fullsolu0}
\end{eqnarray}
where $W(y_0, y_1)(z) = 2/z^2$ is the Wronskian of $y_0$ and $y_1$ and 
\be
\t g_1 := \frac {g_1}{2s\mu^2 \epsilon_0}. 
\ee

Before proceeding some comments are in order: contrary to what happens in the classical commutative case, where we have a homogeneous differential equation for the electric potential~\cite{Vuille:2002qz}, now we have to deal with contributions involving the function $y_1= e^x/x$ that diverges when $x \to \infty$. Hence, what we should do is to guarantee that the solution for $u_0$ is well-behaved in this limit and for that we can choose the constant $c_1$ properly; this is similar to the choice of the constant $b$ in the solution of $e^{\nu_0}$. 

To do this, let us consider the integral in the fourth term in Eq.~(\ref{fullsolu0})
\be
e^{\mu^2\theta} \int_0^x dz \, z e^{-(z + 2 \mu^2\theta)^2/4\mu^2\theta},
\ee
where we have used the explicit expressions for $y_0, y_1$ and $g_1$. We see then that
\be
c_1 = -e^{\mu^2\theta} \int_0^\infty dz \, z e^{-(z + 2 \mu^2\theta)^2/4\mu^2\theta},
\ee
is the appropriate choice that will give a regular solution at infinity due to the fact that the second and fourth term in Eq.~(\ref{fullsolu0}) will then combine to produce upper incomplete gamma functions. When multiplied by the function $y_1$, the resulting function will converge as $x \to \infty$.

\begin{center}
\begin{figure}[bth]
 \includegraphics[width=8cm]{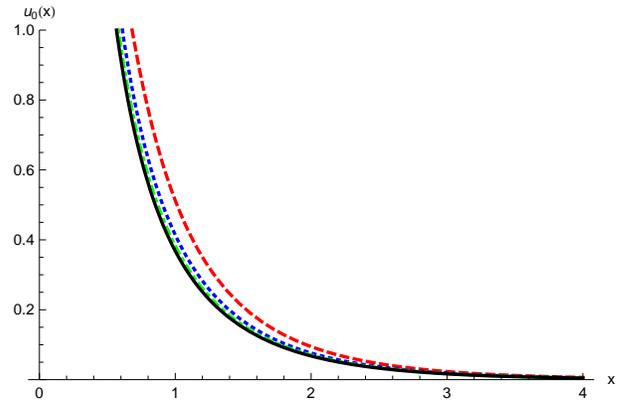}
\caption{The function $\t u_0(x)$ is shown for $a = 0.05, 0.005, 0.0005$ from bottom to top. The solid line is the function $e^{-x}/x$.} 
\label{fig1}
\end{figure}
\end{center}

\begin{center}
\begin{figure}[bth]
\includegraphics[width=8cm]{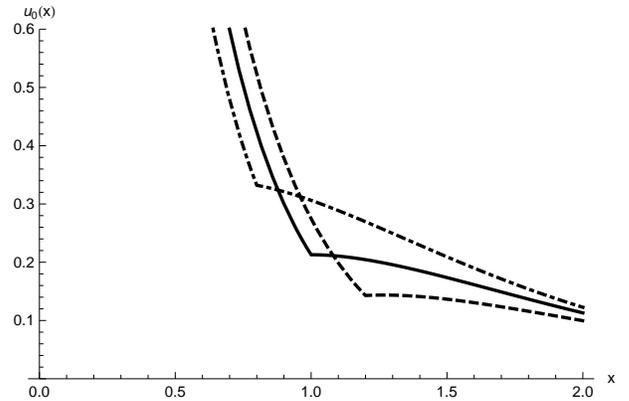}
\caption{The function $\t u_0(x)$ for $a = 0.4, 0.5, 0.6$ from left to right.} 
\label{fig2}
\end{figure}
\end{center}

A similar procedure can be done with the integral in the third term in Eq.~(\ref{fullsolu0}), where after a change of variable $z \to \zeta = z - 2\mu^2\theta$ and a shift of the arbitrary constant $c_0$, the integral can be written now in terms of lower incomplete gamma functions. Taking into account all these considerations, we arrive then finally for $u_0$ to the expression
\be
u_0(x) = c_0 \frac {e^{-x}}x - \frac Q{(4\pi\theta)^{3/2}} \frac {e^{\mu^2\theta}}{4s\mu^2\epsilon_0} \times 2\mu^2\theta \, \t u_0(x)
\label{fu0}
\ee
where
\begin{eqnarray}
&&\t u_0(x) = \frac {e^{-x}}x \left\{ \gamma \left[ 1, \frac {(x - 2 a)^2}{4a} \right] + a^{1/2} \, \gamma \left[ \frac 12, \frac {(x - 2a)^2}{4a} \right] \right\}
\nonumber \\[4pt]
&&+ \frac {e^x}x \left\{ \Gamma \left[ 1, \frac {(x + 2 a)^2}{4a} \right] - a^{1/2} \, \Gamma \left[ \frac 12, \frac {(x + 2a)^2}{4a} \right] \right\},
\label{PNC23}
\end{eqnarray}
$a := \mu^2 \theta$, $\gamma(n, z)$ is the lower incomplete gamma function defined previously and $\Gamma(n, z)$ is the upper incomplete gamma function
\begin{equation}
\Gamma\left( n, z \right) := \int_z^\infty dt\, t^{n-1/2}e^{-t}.
\end{equation}
We obtain thus a very symmetric expression for the dimensionless electric potential $u_0$. In Fig.~\ref{fig1} we plot the function $\t u_0(x)$ for small values of $a$; it can be seen easily that in the limit $a \to 0$ this function becomes $e^{-x}/x$. At the value $x = 2a = 2\mu^2 \theta$, this function has a peak which is more evident for values of $a$ around $0.5$ as shown in Fig.~\ref{fig2}.

Having obtained the zero-order solutions to both the metric and the electromagnetic potential, the constant $c_0$ remains unspecified but it will be fixed later, we now focus on he solutions to first order Eqs.~(\ref{PNC14}). In these equations, we can subtract the second equation to the first to obtain
\begin{equation}
\lambda_1^\prime + \nu_1^\prime = xe^{-2\nu_0}u_0^2,
\label{PNC24}
\end{equation}
and this equation can immediately be solved for $\lambda$, that is 
\begin{equation}
\lambda_1=-\nu_1+\int^x dz\, ze^{-2\nu_0}u_0^2.
\label{PNC25}
\end{equation}
In the same way, solving in Eq.~(\ref{PNC24}) for $\lambda^\prime$ and substituting this and Eq.~(\ref{PNC25}) in the second order differential equation for $\nu_1$, we obtain after some algebra
\begin{eqnarray}
\nu_1^{\prime\prime} + \frac 2x \nu_1^\prime + 2 \nu_1^\prime \nu_0^\prime - \frac {2\kappa}{\mu^2} e^{-\nu_0} h_3 \nu_1
= \frac x2 e^{-2\nu_0} \nu_0^\prime u_0^2 
\nonumber \\[4pt]
+ 2 e^{-2\nu_0}u_0^2 + e^{-\nu_0}{u_0^\prime}^2 - \frac {2\kappa}{\mu^2} e^{-\nu_0}h_3\int^x dz \, z e^{-2\nu_0} u_0^2.
\label{PNC26}
\end{eqnarray}

To simplify this equation further, we may take into account now the same considerations that we discussed in the case of the zero order equations. The basic idea is to recall that the Schwarzschild-like term in $e^{\nu_0}$ will define a small function 
\be
\epsilon(x) = -  \frac {\kappa\mu M}{4\pi^{3/2}} \frac 1x \gamma \left( \frac 32, \frac {x^2}{4a} \right).
\ee
More precisely, the parameter
\be
\delta := \frac {\kappa\mu M}{8\pi^{3/2}},
\ee
will be taken to be small. In consequence we have the following expressions 
\begin{equation}
e^{-\nu_0}=1-\epsilon(x), \qquad e^{-2\nu_0}=1-2\epsilon(x),
\label{PNC28}
\end{equation}
together with
\begin{equation}
\nu_0^\prime = \epsilon^\prime(x).
\label{PNC27}
\end{equation}

Now, since by definition $\nu_1(x)$ is a first order correction as well as its derivative, they can be assumed to be both of order on $\delta$ and accordingly we can eliminate higher order terms in the left hand side of Eq.~(\ref{PNC26}). By doing so, we arrive to the following differential equation for $\nu_1$ 
\be
\nu_1^{\prime\prime} + 2 \frac {\nu_1^\prime}x = F(x),
\label{eq4nu1}
\ee
where $F(x)$ is defined as the right hand side of Eq.~(\ref{PNC26}) after simplification. Once this function is computed, we obtain therefore the solution
\be
\nu_1 =  d_0 + \frac {d_1}x + \int_0^x dz \, z F(z) - \frac 1x \int_0^x dz \, z^2 F(z).
\label{solnu1}
\ee
Finally, with this input the metric coefficient $g_{00}$ will be given by
\be
g_{00} =  e^\nu = 1 -  \frac {\kappa\mu M}{2\pi^{3/2}} \frac 1x \gamma \left( \frac 32, \frac {x^2}{4a} \right) + \nu_1(x),
\label{finalsolg00}
\ee
in our approximation.       

\section{Asymptotic limits and regularity}
\label{secc:4}

Before proceeding and as a check to our calculations, it is worthwhile to take a moment to discuss the limits $\mu \neq 0, \theta \to 0$ and $\mu \to 0, \theta \neq 0$. In the first case, the solution to the scaled electromagnetic potential is  
\begin{equation}
u_0(x)=c_0 \frac {e^{-x}}x.
\end{equation} 
and the corresponding equation equation for the perturbation $\nu_1(x)$ becomes 
\begin{equation}
\nu_1^{\prime\prime} (x) + \frac 2x \nu_1^\prime (x) = u_0^\prime (x)^{2} + 2u_0(x)^{2}.
\end{equation}
Using the explicit expression for $u_0(x)$ in this last equation we obtain
\begin{equation}
\nu_1^{\prime\prime} (x) + \frac 2x \nu_1^\prime (x) = c_0^2\biggl( 3{e^{-2x}\over x^2} + 2{e^{-2x}\over x^3} + {e^{-2x}\over x^4}\biggl ).
\end{equation}
As expected, we recover the differential equation for $\nu_1$ discussed in~\cite{Vuille:2002qz}. The particular solution to it is then
\begin{equation}
\nu_1(x)={c_0^2\over 2}\biggl({e^{-2x}\over x^2}-\int {e^{-2x}\over x^2}dx\biggl).
\end{equation}

For the second case, the analysis is better done starting with the differential equation Eq.~(\ref{PNC22}) for $u_0$ and using the radial variable $r$. With the explicit form of the function $g_1$ we have
\begin{equation}
u_0^{\prime\prime}(r) + {2\over r}u_0^\prime(r) = {1\over 2s\epsilon_0} {Q\over (4\pi\theta)^{3/2}}e^{-r^2/4\theta},
\end{equation}
or equivalently
\be
(r^2 u_0^\prime(r))^\prime = \frac 1{2s\epsilon_0} {Q\over (4\pi\theta)^{3/2}} \, r^2 e^{-r^2/4\theta}.
\ee
Integration gives immediately
\be
u_0^\prime(r) = \frac 1{2s\epsilon_0} {Q\over (4\pi\theta)^{3/2}} \frac 1{r^2} \int_0^r \, dz \, z^2 e^{-z^2/4\theta}.
\ee
After a change of variables in the integrand, we arrive to
\be
u_0^\prime(r) = \frac 1{4s\epsilon_0} \frac Q{\pi^{3/2}} \frac 1{r^2} \gamma \left( \frac 32, \frac {r^2}{4\theta} \right).
\label{deru0}
\ee

The previous result can also be derived from the solution
\begin{equation}\label{PNC1.1}
u_0(r)=-{2\theta A\over r}\int e^{-r^2/4\theta}dr,
\end{equation}
obtained from the variation of parameters method. Here $A := Q/ 2s\epsilon_0(4\pi\theta)^{3/2}$. From it we have
\begin{equation}
u^\prime_0(r) = {2\theta A\over r^2}\int e^{-r^2/4\theta}dr-{2\theta A\over r}e^{-r^2/4\theta}.
\end{equation}
After an integration by parts we finally arrive to  
\begin{equation}
u^\prime_0(r) = {A\over r^2}\int r^2e^{-r^2/4\theta}dr,
\end{equation} 
which is nothing but Eq.~(\ref{deru0}). Thus, the electric field is given by
\be
E(r) = -\frac Q{4\pi\epsilon_0 r^2} \frac 1{\pi^{1/2}} \gamma \left( \frac 32, \frac {r^2}{4\theta} \right),
\ee
which is, up to a constant factor, the same expression as that obtained by Nicolini~\cite{Nicolini:2008aj}.

Having discussed these two limits, we now want to consider solutions to Eq.~(\ref{PNC26}) such that {\it both} the scaled magnetic potential $u_0$ and the metric function $\nu_1$ be regular at the origin. This requirement for the function $u_0$ fixes the value of the constant $c_0$ in Eq.~(\ref{fu0}) to 
\be
c_0 = \sqrt\pi \chi e^a + \frac {\chi e^a}{\sqrt{a}} -2 \chi e^a \Gamma \left( \frac 12, a \right),
\label{c0value}
\ee
where $\chi := Q/4\pi^{1/2}q$. Near $x = 0$ we have then the following series development
\be
\chi^{-1} u_0 (x) = 2e^a \Gamma \left( \frac 12, a \right) - \frac 1{\sqrt{a}} x + \dots
\ee
The parameter $\chi$ measures the relative strength of the charge $Q$ of a test particle to the Proca charge $q$; we may assume that $\chi \ll 1$. Then from the second and third terms in the right hand side of Eq.~(\ref{PNC26}) we notice that $\chi^2 \sim \delta$. This in turn means that the integral in this equation can be neglected, its contribution being of order $\delta \chi^2 \sim \delta^2$ and furthermore, than the first term involving $\nu_0^\prime u_0^2 \sim \delta^2$ can also be omitted.

In this reduced setting the function $F(x)$ is given then by
\be
F(x) = {u_0^\prime}^2 + 2 u_0^2, 
\ee
where we have used $e^{\nu_0} = 1$ to lowest order. Around $x = 0 $ it has the following behaviour
\be
F(x) = \frac {\chi^2}a \left[ 1 + 8 e^a a \Gamma \left( \frac 12, a \right)^2 \right] - \frac {28 e^a \chi^2 \Gamma \left( \frac 12, a \right)}{3\sqrt{a}} x + \dots,
\ee
being then regular at the origin. Recalling that $u_0$ vanishes at infinity, then $u^\prime_0 \to 0$ as $x \to \infty$ and hence, $F(x)$ also vanishes as $x \to \infty$.

It is clear now that by taking $d_0 = d_1 = 0$ in Eq.~(\ref{solnu1}) we have a regular solution for $\nu_1(x)$ at $x= 0$. Indeed
\be
\nu_1(x) = \frac {\chi^2}a \left[1 + 8 e^a a \Gamma \left( \frac 12, a \right)\right] x + \dots
\ee
when $x \ll1$. 

On the other hand, from the fact that the function $F(x)$ is positive definite, and finite at $x = 0$, we see that the derivative
\be
\nu_1^\prime = \frac 1{x^2} \int_0^x z^2 F(z)
\ee
will always be positive; the integral converges since $F(x)$ vanishes exponentially for large values of $x$. Inflection points for $\nu_1$ will satisfy the equation
\be
x^3 F(x) = 2 \int_0^x z^2 F(z). 
\ee
Therefore, since $\nu_1$ vanishes at the origin and its derivative is always positive, we deduce that $\nu_1$ is always positive. In consequence any horizon arising from the zero order solution Eq.~(\ref{PNC17}) will be shifted from its original location, and indeed, a horizon in the zero order solution may be eliminated when taking into account the correction $\nu_1$.

\subsection{Strong non-commutativity}
\label{secc:5}

\begin{center}
\begin{figure}[bth]
\includegraphics[width=8cm]{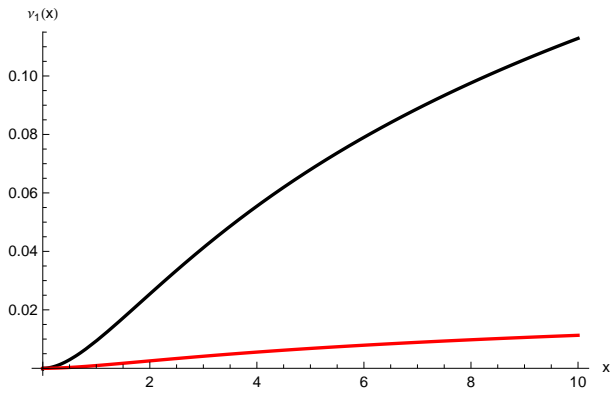}
\caption{The function $\nu_1(x)$ for $a = 1, 10$ ($\chi=0.1$) from top to bottom.} 
\label{fig3}
\end{figure}
\end{center}

For the particular choice of $c_0$ given by Eq.~(\ref{c0value}) in the previous section, let us now consider the case $a \gg 1$, which is the regime of strong non-commutativity. We focus on it because explicit calculations can be done. First we have the asymptotic expression
\be
u_0(x) = \frac {2\chi}{\sqrt{a}} \frac {1 - e^{-x}}x = \frac {2\chi}{\sqrt{a}} 
\left\{ \begin{matrix}
1 - \frac 12 x & \mbox { for } x \ll 1, \\[4pt]
\frac 1x & \mbox { for } 1 \ll x, 
\end{matrix} \right.
\ee
as $a \to \infty$. The function $F(x)$ in this case is given by 
\begin{eqnarray}
F(x) &=& \frac {4\chi^2}{a x^4} \left[ 1 + 2x^2 - 2e^{-x} - 2 x e^{-x} - 4 x^2 e^{-x} \right.
\nonumber \\[4pt]
&&\left. +e^{-2x} + 2 x e^{-2x} +3 x^2 e^{-2x} \right].
\end{eqnarray}
The two integrals $\int_0^x dz \, z F(z)$ and $\int_0^x dz \, z^2 F(z)$ can be readily calculated. We obtain
\be
\nu_1(x) = \frac {\chi^2}a [4f_1(x) - 2 \frac {e^{-2x}}{x^2} f_2(x)],
\ee
where 
\begin{eqnarray}
f_1(x) &=& f_0 - \frac 1{2x^2} [1 - 2e^{-x} - 2 x e^{-x} + e^{-2x} + 2 x e^{-2x}
\nonumber \\[4pt]
&& - 2x^2 Ei(-2x) + 6x^2 Ei(-x) - 4 x^2 \ln (x) ],
\end{eqnarray}
and
\be
f_2(x) = -2 -3x + 4 e^x + 8 x e^x - 2 e^{2x} -5x e^{2x} + 4x^2 e^{2x}.
\ee
Here $f_0 := -\frac 12 + 2\gamma - \ln(2)$, $\gamma$ is Euler's constant and
\be
Ei(z) = \int_{-z}^\infty \frac {e^{-t}}t dt.
\ee
In Fig.~\ref{fig3} we plot this solution for some values of the parameter $a$. We remark that it has the behaviour discussed previously.

\section{Conclusions}

We have analysed a non-commutative gravitational model based on the non-linear Proca-Einstein Lagrangian. The presence of non-commutative relations among the coordinates is implemented by considering smeared mass and charge distributions. 

It has been shown that the field equations can be solved perturbatively by following a procedure similar to that of the classical case, namely by doing a series development on the Proca mass parameter $\mu$. This allowed us to write a set of coupled equations for the zero order metric functions and the corresponding first order corrections. The solution of the first set gives as expected for vanishing non-commutative parameter the classical solution of Vuielle et al.~\cite{Vuille:2002qz}, while in the limit of small Proca mass, the non-commutative solution for a spherical space-time given by Nicolini~\cite{Nicolini:2008aj} is recovered. 

Among the possible solutions to the field equations we have considered a certain class defined by having regular metric functions both at infinity and the origin. That this kind of solution exists is not {\it a priori} evident since now the differential equation for the scaled potential $u_0$ is modified in a non-trivial way, its general solution given in terms of decreasing and raising exponentials and lower and upper incomplete gamma functions. 

Fortunately enough, it is possible to choose the integration constants appropriately to satisfy the requirements of regularity.  An important feature of this class of solutions is the fact that horizons may be avoided by the contribution coming from the correction $\nu_1$ and this holds as long $a \neq 0$; non-commutativity, in spite of producing a complicated set of field equations, is essential in this regard. 

Finally, in the limit of strong non-commutativity, as measured by the parameter $a \to \infty$, we observe that the scaled potential $u_0$ for the Einstein-Proca system has a very simple form and explicit calculations can then be performed. We see that $\nu_1$ takes small values in this regime so that it can be seen as a small perturbation to the zero order metric coefficient $e^{\nu_0}$ which behaves as
\be
1 - \frac 13 \frac \delta{a^{3/2}} x,
\ee
when $a\to \infty$. We may try to apply these and related ideas to other kinds of non-linear electrodynamics.

\acknowledgments
This research was supported by CONACyT-DFG Collaboration Grant 147492 ``Noncommutative Models in Physics". O. S.-S. was also supported by a Posdoctoral Fellowship Grant PROMEP/103.5/13/9043.


\begin{thebibliography}{17}
\expandafter\ifx\csname natexlab\endcsname\relax\def\natexlab#1{#1}\fi
\expandafter\ifx\csname bibnamefont\endcsname\relax
  \def\bibnamefont#1{#1}\fi
\expandafter\ifx\csname bibfnamefont\endcsname\relax
  \def\bibfnamefont#1{#1}\fi
\expandafter\ifx\csname citenamefont\endcsname\relax
  \def\citenamefont#1{#1}\fi
\expandafter\ifx\csname url\endcsname\relax
  \def\url#1{\texttt{#1}}\fi
\expandafter\ifx\csname urlprefix\endcsname\relax\def\urlprefix{URL }\fi
\providecommand{\bibinfo}[2]{#2}
\providecommand{\eprint}[2][]{\url{#2}}

\bibitem[{\citenamefont{Groenewold}(1946)}]{Groenewold:1946kp}
\bibinfo{author}{\bibfnamefont{H.~J.} \bibnamefont{Groenewold}},
  \bibinfo{journal}{Physica} \textbf{\bibinfo{volume}{12}},
  \bibinfo{pages}{405} (\bibinfo{year}{1946}).

\bibitem[{\citenamefont{Moyal}(1949)}]{Moyal:1949sk}
\bibinfo{author}{\bibfnamefont{J.~E.} \bibnamefont{Moyal}},
  \bibinfo{journal}{Proc. Cambridge Phil. Soc.} \textbf{\bibinfo{volume}{45}},
  \bibinfo{pages}{99} (\bibinfo{year}{1949}).

\bibitem[{\citenamefont{Snyder}(1947{\natexlab{a}})}]{Snyder:1946qz}
\bibinfo{author}{\bibfnamefont{H.~S.} \bibnamefont{Snyder}},
  \bibinfo{journal}{Phys. Rev.} \textbf{\bibinfo{volume}{71}},
  \bibinfo{pages}{38} (\bibinfo{year}{1947}{\natexlab{a}}).

\bibitem[{\citenamefont{Snyder}(1947{\natexlab{b}})}]{Snyder:1947nq}
\bibinfo{author}{\bibfnamefont{H.~S.} \bibnamefont{Snyder}},
  \bibinfo{journal}{Phys. Rev.} \textbf{\bibinfo{volume}{72}},
  \bibinfo{pages}{68} (\bibinfo{year}{1947}{\natexlab{b}}).

\bibitem[{\citenamefont{Smailagic and Spallucci}(2003)}]{Smailagic:2003rp}
\bibinfo{author}{\bibfnamefont{A.}~\bibnamefont{Smailagic}} \bibnamefont{and}
  \bibinfo{author}{\bibfnamefont{E.}~\bibnamefont{Spallucci}},
  \bibinfo{journal}{J.Phys.} \textbf{\bibinfo{volume}{A36}},
  \bibinfo{pages}{L517} (\bibinfo{year}{2003}), \eprint{hep-th/0308193}.

\bibitem[{\citenamefont{Smailagic and Spallucci}(2004)}]{Smailagic:2004yy}
\bibinfo{author}{\bibfnamefont{A.}~\bibnamefont{Smailagic}} \bibnamefont{and}
  \bibinfo{author}{\bibfnamefont{E.}~\bibnamefont{Spallucci}},
  \bibinfo{journal}{J.Phys.} \textbf{\bibinfo{volume}{A37}}, \bibinfo{pages}{1}
  (\bibinfo{year}{2004}), \eprint{hep-th/0406174}.

\bibitem[{\citenamefont{Nicolini et~al.}(2006)\citenamefont{Nicolini,
  Smailagic, and Spallucci}}]{Nicolini:2005vd}
\bibinfo{author}{\bibfnamefont{P.}~\bibnamefont{Nicolini}},
  \bibinfo{author}{\bibfnamefont{A.}~\bibnamefont{Smailagic}},
  \bibnamefont{and}
  \bibinfo{author}{\bibfnamefont{E.}~\bibnamefont{Spallucci}},
  \bibinfo{journal}{Phys. Lett.} \textbf{\bibinfo{volume}{B632}},
  \bibinfo{pages}{547} (\bibinfo{year}{2006}), \eprint{arXiv: gr-qc/0510112}.

\bibitem[{\citenamefont{Nicolini}(2009)}]{Nicolini:2008aj}
\bibinfo{author}{\bibfnamefont{P.}~\bibnamefont{Nicolini}},
  \bibinfo{journal}{Int. J. Mod. Phys.} \textbf{\bibinfo{volume}{A24}},
  \bibinfo{pages}{1229} (\bibinfo{year}{2009}), \eprint{arXiv: 0807.1939
  [hep-th]}.

\bibitem[{\citenamefont{Tejeiro and Larranaga}(2012)}]{Tejeiro:2010gu}
\bibinfo{author}{\bibfnamefont{J.~M.} \bibnamefont{Tejeiro}} \bibnamefont{and}
  \bibinfo{author}{\bibfnamefont{A.}~\bibnamefont{Larranaga}},
  \bibinfo{journal}{Pramana} \textbf{\bibinfo{volume}{78}},
  \bibinfo{pages}{155} (\bibinfo{year}{2012}), \eprint{1004.1120}.

\bibitem[{\citenamefont{Liang and Liu}(2012)}]{Liang:2012vx}
\bibinfo{author}{\bibfnamefont{J.}~\bibnamefont{Liang}} \bibnamefont{and}
  \bibinfo{author}{\bibfnamefont{B.}~\bibnamefont{Liu}},
  \bibinfo{journal}{Europhys. Lett.} \textbf{\bibinfo{volume}{100}},
  \bibinfo{pages}{30001} (\bibinfo{year}{2012}).

\bibitem[{\citenamefont{Rahaman et~al.}(2013)\citenamefont{Rahaman, Kuhfittig,
  Bhui, Rahaman, Ray et~al.}}]{Rahaman:2013gw}
\bibinfo{author}{\bibfnamefont{F.}~\bibnamefont{Rahaman}},
  \bibinfo{author}{\bibfnamefont{P.}~\bibnamefont{Kuhfittig}},
  \bibinfo{author}{\bibfnamefont{B.}~\bibnamefont{Bhui}},
  \bibinfo{author}{\bibfnamefont{M.}~\bibnamefont{Rahaman}},
  \bibinfo{author}{\bibfnamefont{S.}~\bibnamefont{Ray}}, \bibnamefont{et~al.},
  \bibinfo{journal}{Phys. Rev.} \textbf{\bibinfo{volume}{D87}},
  \bibinfo{pages}{084014} (\bibinfo{year}{2013}), \eprint{1301.4217}.

\bibitem[{\citenamefont{Proca}(1936)}]{Proca:1936}
\bibinfo{author}{\bibfnamefont{A.}~\bibnamefont{Proca}}, \bibinfo{journal}{J.
  Phys. Radium} \textbf{\bibinfo{volume}{7}}, \bibinfo{pages}{347}
  (\bibinfo{year}{1936}).

\bibitem[{\citenamefont{Pauli}(1941)}]{Pauli:1941zz}
\bibinfo{author}{\bibfnamefont{W.}~\bibnamefont{Pauli}}, \bibinfo{journal}{Rev.
  Mod. Phys.} \textbf{\bibinfo{volume}{13}}, \bibinfo{pages}{203}
  (\bibinfo{year}{1941}).

\bibitem[{\citenamefont{Tu et~al.}(2005)\citenamefont{Tu, Luo, and
  Gillies}}]{Tu:2005ge}
\bibinfo{author}{\bibfnamefont{L.-C.} \bibnamefont{Tu}},
  \bibinfo{author}{\bibfnamefont{J.}~\bibnamefont{Luo}}, \bibnamefont{and}
  \bibinfo{author}{\bibfnamefont{G.}~\bibnamefont{Gillies}},
  \bibinfo{journal}{Rept. Prog. Phys.} \textbf{\bibinfo{volume}{68}},
  \bibinfo{pages}{77} (\bibinfo{year}{2005}).

\bibitem[{\citenamefont{Casana et~al.}(2008)\citenamefont{Casana, Ferreira, and
  Santos}}]{Casana:2008nw}
\bibinfo{author}{\bibfnamefont{R.}~\bibnamefont{Casana}},
  \bibinfo{author}{\bibfnamefont{J.}~\bibnamefont{Ferreira},
  \bibfnamefont{Manoel~M.}}, \bibnamefont{and}
  \bibinfo{author}{\bibfnamefont{C.~E.} \bibnamefont{Santos}},
  \bibinfo{journal}{Phys. Rev.} \textbf{\bibinfo{volume}{D78}},
  \bibinfo{pages}{025030} (\bibinfo{year}{2008}), \eprint{0804.0431}.

\bibitem[{\citenamefont{Cheon et~al.}(2009)\citenamefont{Cheon, Lee, and
  Lee}}]{Cheon:2009zx}
\bibinfo{author}{\bibfnamefont{S.}~\bibnamefont{Cheon}},
  \bibinfo{author}{\bibfnamefont{C.}~\bibnamefont{Lee}}, \bibnamefont{and}
  \bibinfo{author}{\bibfnamefont{S.~J.} \bibnamefont{Lee}},
  \bibinfo{journal}{Phys. Lett.} \textbf{\bibinfo{volume}{B679}},
  \bibinfo{pages}{73} (\bibinfo{year}{2009}), \eprint{0904.2065}.

\bibitem[{\citenamefont{Vuille et~al.}(2002)\citenamefont{Vuille, Ipser, and
  Gallagher}}]{Vuille:2002qz}
\bibinfo{author}{\bibfnamefont{C.}~\bibnamefont{Vuille}},
  \bibinfo{author}{\bibfnamefont{J.}~\bibnamefont{Ipser}}, \bibnamefont{and}
  \bibinfo{author}{\bibfnamefont{J.}~\bibnamefont{Gallagher}},
  \bibinfo{journal}{Gen. Rel. Grav.} \textbf{\bibinfo{volume}{34}},
  \bibinfo{pages}{689} (\bibinfo{year}{2002}).

\end{thebibliography}

\end{document}